\documentclass{WileyMSP-template}
\usepackage{siunitx}
\usepackage{amsmath}
\DeclareSIUnit{\litre}{l}
\DeclareSIUnit\Molar{M}
\usepackage{textcomp}
\usepackage{gensymb}
\usepackage{physunits}
\usepackage{cite}
\usepackage{xcolor}
\usepackage{hyperref}
\usepackage{xr-hyper} 
\DeclareSIUnit\bar{bar}
\usepackage[version=4]{mhchem}
\definecolor{linkcolor}{rgb}{0.5,0.1,0.1}
\definecolor{urlcolor} {rgb}{0.1,0.1,0.5}
\definecolor{citecolor}{rgb}{0.1,0.5,0.1}
\hypersetup{colorlinks=true,linkcolor=linkcolor,urlcolor=urlcolor,citecolor=citecolor,pdfborder={0 0 0}}
\RenewCommandCopy\qty\SI

\begin{document}

\pagestyle{fancy}
\newcommand{\fixme}[1]{{\color{red} [[#1]]}}
\rhead{}

\title{Self-Assembly of Lipid-Biopolymer Periodic Nanostructures on Photonic Length Scales}

\maketitle

\author{Rushna Quddus*}
\author{Meron Debas}
\author{Stefan Salentinig}
\author{Ullrich Steiner}
\author{Viola Vogler-Neuling*}

\begin{affiliations}
R.Q., U.S. and V.V.N.\\
Adolphe Merkle Institute, Fribourg, Switzerland\\
National Center for Competence in Research, Bioinspired Materials, University of Fribourg, Fribourg, Switzerland\\
viola.vogler-neuling@unifr.ch; rushna.quddus@unifr.ch

M.D. and S.S.\\
Department of Chemistry, University of Fribourg, Fribourg, Switzerland\\
National Center for Competence in Research, Bioinspired Materials, University of Fribourg, Fribourg, Switzerland
\end{affiliations}


\begin{abstract}
The self-assembly of photonic nanostructures in insects involves chitin, proteins, and lipids. 
While synthetic photonic systems have been extensively studied, current lipid-based self-assembly systems are limited in periodicity to $\SI{68}{\nano\meter}$ compared to photonic length scales ($\approx \SI{450}{\nano\meter}$) observed in biological organisms. We hypothesise that lipids facilitate how structural colour arises in vivo by acting as templates for the self-assembly of biopolymers via lipidic lyotropic liquid crystal mesophases. 
Here, we aim to understand and identify how structural colour is produced in insects by the co-assembly of lipids and biopolymers. We study the effect of biopolymers, pH, temperature, surface charge, and stability on lipid vesicles using dynamic light scattering, X-ray scattering, and zeta potential analysis. Using cryo-electron microscopy, we demonstrate that these vesicles interact with the biopolymers and generate periodic nanostructures with periodicities ranging from $\SI{700}{\nano\meter}$ to $\SI{1.2}{\micro\meter}$ more than ten times larger than for purely lipidic systems and dimensionalities ranging from 1D to 3D. Our results establish that lipid mesophases and biopolymers can induce reorganisation into ordered nanostructures, overcoming key limitations of periodicities achieved by lipid-only systems, and providing a methodology for recreating the physicochemical mechanisms underlying biophotonic structural colour.
\end{abstract}

\section{Introduction}

\begin{figure}
\centering
\includegraphics[width=1\linewidth]{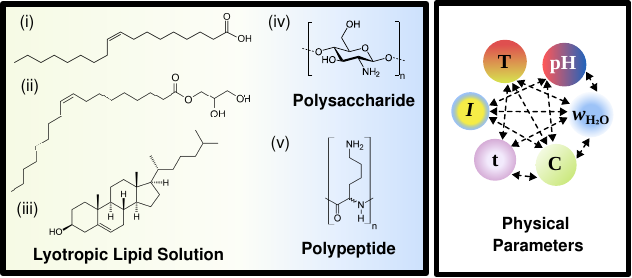}
\caption{ 
\textbf{Explored parameter space with the aim of creating photonic, self-assembled lipidic morphologies.}
The examined solutions comprise biopolymers (polysaccharides and polypeptides), as well as a pair of lyotropic lipids. Representative chemicals used are: (i) oleic acid (ii) monoolein (iii) cholesterol, (iv) chitosan and (v) poly-(L-lysine). The following parameters can be optimised to facilitate the self-assembly of periodic nanostructures: pH, lipid and biopolymer concentrations ($C$), hydration ($w_{\ce{H2O}}$), ionic strength ($I$), temperature ($T$) and time ($t$).
}
\label{fig1}
\end{figure}

Many organisms synthesise cuticular morphologies that exhibit characteristic structures on a 100-nm length scale \cite{mezzenga2019nature}. There ordered or partially disordered structures give rise to ``structural colour'' in these organisms. 
These photonic architectures are prevalent in arthropods, such as butterflies, beetles, spiders, and crustaceans, as well as in birds, where they play a crucial role in camouflage, communication, and attraction \cite{Barrows2014}.  While the optical mechanisms of a limited number of species have been partially elucidated \cite{Mcdougal2021, Ghiradella2000}, the vast majority of insect photonic structures, as well as the mechanisms that lead to their formation within butterfly pupae, remain elusive. This knowledge gap hinders the biomimetic recreation of these architectures.

Several hypotheses have been put forward to explain the formation process. These range from the folding of the endoplasmic reticulum and the self-assembly of lipid-bilayer membranes \cite{saranathan2010structure} to mechanisms involving spinodal decomposition \cite{zhang2013replication}. Early studies suggested that complex nanomorphologies arise from the self-assembly of lipid membranes, a concept supported by detailed observations of wing scale growth \cite{Ghiradella1989}. Another hypothesis is that these photonic nanostructures assemble within the epithelial cells of the developing wing scales. This view is supported by recent meta-analyses of butterfly structural colouration, which emphasise the importance of organising the chitin-lipid/protein matrix \cite{stossberg1938zellvorgange, Thayer2023}. Subsequent research has employed advanced microscopy, super-resolution techniques to visualise these formation processes \cite{Dinwiddie.2014, Mcdougal2021, SeahSaranathan2023}. These nanostructures are difficult to observe \textit{in vivo} because they are too small to be detected by standard light microscopy and electron microscopy does not allow for time-dependent observation.

In this study, we adopt a bottom-up approach to self-assembly to determine if photonic crystals can be produced artificially using self-assembled lipidic lyotropic liquid crystals (LLLC) \cite{Mezzenga2019}. This approach ensures that any resulting periodic nanostructured materials are inherently biocompatible and environmentally sustainable. This study explores the parameter space shown in Figure \ref{fig1} with the aim of establishing a methodology for the self-organisation of lipid structures on a 100\,nm length scale. The objective is to replicate the processes nature uses to generate photonic morphologies.

Periodic nanostructures in nature exhibit various types of symmetry, including lamellar ($L$), inverse hexagonal $(H_{\rm II})$, gyroid $(Ia\overline{3}d)$ and diamond $(Pn\overline{3}m)$. More intricate structures, such as I-WP, have also been identified \cite{Vogler-Neuling2023}. 
LLLCs are a self-assembly system capable of generating these symmetries from biological components \cite{mezzenga2019nature}. 
Lipidomic analyses of butterflies such as \textit{Morpho peleides}, \textit{Danaus plexippus} and \textit{Papilio polytes} have identified oleic acid as a ubiquitous molecule \cite{Wang2006,Reisinger1969,mura2022}. 

The mixture of oleic acid and monoolein exhibits several lyotropic liquid crystal phases \cite{Salentinig2010, Nakano2002}. 
Although inverse bicontinuous lipidic liquid crystalline phases, such as the cubic $Im\overline{3}m$, can form nanostructures with large unit cells, the largest achieved for membrane protein crystallisation via electrostatic swelling is $\SI{68}{\nano\meter}$ \cite{Kim2017}. 
This suggests that, while highly symmetric mesostructures can be formed with simple molecular compositions, a significant effort is required to understand how insects achieve the periodicities of several hundred nanometres -- necessary for visible-light photonic responses -- using a truly biomimetic approach.

To address this issue, we explore the parameter space illustrated in Figure \ref{fig1}. This approach investigates the structure-function relationships that govern the self-assembly of photonic crystals in insects by systematically varying various compositional and physical parameters.
These parameters include the molecular composition, which can include relevant lipids and biopolymers for studying structural colour in living organisms, primarily lipids and cholesterol, polypeptides and proteins, and polysaccharides and sugars. Additionally, the self-assembly parameters encompass physical properties that can affect the system, such as pH, hydration ($w_{\ce{H2O}}$), concentration ($C$), ionic strength ($I$), temperature ($T$), and time ($t$) to examine the self-assembly of periodic nanostructures.
Figure~\ref{fig1} illustrates a schematic for the explored parameter space with the goal to recreate photonic crystals seen in living organisms.

To replicate these nanostructures, we selected lyotropic lipids \cite{Mezzenga2019}, polypeptides \cite{Dinwiddie.2014} and chitosan  \cite{peter1986structural}. We selected chitosan because the nanostructures in insects are hypothesised to be made of chitinous cuticle, and chitosan is a water-soluble substitute for chitin that cannot yet be produced using enzymes. 
As the photonic crystal development is highly complex, we focus in this work on testing the following three hypotheses in the field to achieve photonic nanostructures from chitin/lipid systems with periodicities on the order of $\approx \SI{450}{\nano\meter}$

\begin{enumerate} 
\item A lipidic liquid crystal phase can localise within a chitinous network and restructure it. 
\item Cholesterol in lipid phases increases the stiffness and decreases curvature in bicontinuous membrane structures, hereby leading to larger periodicities. 
\item Cholesterol increases membrane thickness.
\item Physical parameters that fail to yield photonic crystals artificially are unlikely to be part of the natural formation process. 
\end{enumerate} 
In general, our methodology allows us to test different (physical) parameters and correlating the molecular and physical requirements for periodic nanostructures. Parameters that fail to yield photonic crystals artificially may be unlikely to be part of the natural formation process.
This systematic investigation aims to establish a methodology for creating periodic nanostructures with the dimensions necessary for a photonic response in the visible wavelength range.

\section{Results}

\subsection{Cholesterol-induced changes in lipid membranes do not result in structures at photonic length scales}

To investigate the hypothesised role of lipid composition in photonic structure assembly, lipid vesicle mesophases were selected for analysis. 
This is because they can be easily characterised using well-established methods, making them an ideal starting point for exploring photonic structure formation in nature. 

To evaluate the hypothesis that cholesterol acts as a curvature-reducing agent capable of expanding lipid mesophases to photonic dimensions, the cholesterol concentration in oleic acid/monoolein (OA/MO) vesicles was systematically varied from $\SI{10} {\percent}$ to $\SI{40} {\percent}$ and the corresponding diameter changes were studied. 

\begin{figure}
\centering
\includegraphics[width=1\linewidth]{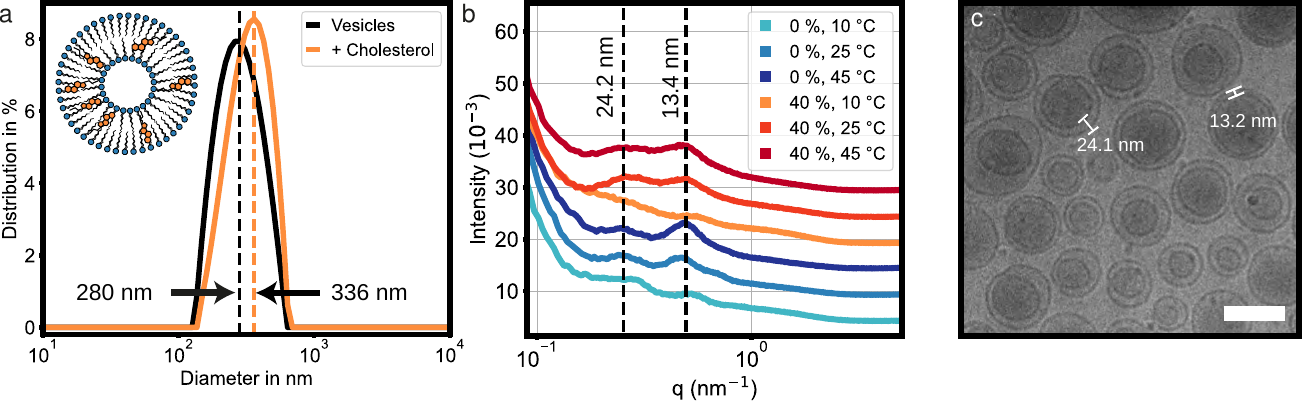}
\caption{\textbf{Characterisation of extruded lipid vesicles.} (a) Change in hydrodynamic diameter when $\SI{30} {\percent}$ cholesterol was added to lyotropic oleic acid -- monoolein ($\SI{30} {\percent}$ OA, $\SI{70} {\percent}$ MO) vesicles extruded in a 30 mM PBS buffer at pH 8.2. (b) SAXS scans of vesicles containing $\SI{0} {\percent}$ and $\SI{40} {\percent}$ cholesterol were performed at $\SI{10}{\celsius}$, $\SI{25}{\celsius}$ and $\SI{45}{\celsius}$ to determine membrane thickness and stability. (c) Cryogenic electron microscope image of extruded vesicles. Scale bar: 200\,nm.
\label{fig2}}
\end{figure}

\subsubsection{Lipid vesicle extrusion}
Lyotropic lipid vesicles composed of oleic acid/monoolein (OA/MO) were produced by extrusion. 
The lipid solutions were dispersed in a phosphate-buffered saline (PBS) solution and then extruded through membranes with pore sizes of 600\,nm, 400\,nm and 100\,nm using a syringe-based extruder, see supplementary Figure S1. 
The formation of the vesicles was confirmed by dynamic light scattering (DLS), small angle X-ray scattering (SAXS) and cryo-transmission electron microscopy (Cryo-TEM), as shown in Figure~\ref{fig2}. 
Lipid vesicles composed of oleic acid and monoolein (30:70) were extruded using a membrane with a pore size of 600\,nm at $\SI{45}{\celsius}$, with the lipid dispersion prepared by thin-film hydration. 
DLS was used to  determine the hydrodynamic radius to be $d = \SI {280}{\nano\meter}$ with a polydispersity index (PDI) of 0.205. 

\subsubsection{Effect of cholesterol on lipid vesicle size}

Lyotropic lipid vesicles composed of OA/MO (30:70) were extruded through 600 nm membranes at $\SI {45}\celsius$ using lipid solutions containing various amounts of cholesterol ranging from $\SI{10} {\percent}$ to $\SI{40} {\percent}$. 
DLS analysis revealed that the addition of $\SI{33}{\percent}$ cholesterol increased the hydrodynamic diameter from $\SI {280}{\nano\meter}$ (PDI 0.205) to $\SI {336}{\nano\meter}$ (PDI 0.218). 
While this represents an increase of approximately $\SI {56}{\nano\meter}$, it falls within the margin of error given an average PDI of 0.2. Therefore, it can be concluded that cholesterol does not significantly alter the vesicle size under extrusion conditions (Figure \ref{fig2}a) and cannot be responsible for a size increase up to hundreds of nanometers.

\subsubsection{Effect of cholesterol on lipid membrane strructure}

SAXS was employed to analyse the structure and stability of the lipid membranes at different temperatures ($\SI{10}\celsius$, $\SI{25}\celsius$, $\SI{45}\celsius$) and cholesterol concentrations (0 \%, 40 \%). 
Membrane stability was evaluated by heating the lipid dispersion to $\SI{45}{\celsius}$ (i.e., above its melting point) and then cooling to $\SI{10}{\celsius}$.
The characteristic scattering peaks at $q = \SI{0.46}{\nano\meter^{-1}}$ and $q = \SI{0.26}{\nano\meter^{-1}}$ remained consistent across all conditions, indicating that membranes remained stable regardless of the amount of cholesterol present (Figure \ref{fig2}b). 
The corresponding $d$-spacings, the center-to-center distance between parallel layers in a periodic structure, of $\SI {13.4}{\nano\meter}$ and $\SI {24.2}{\nano\meter}$ are attributed to multi-lamellar ordering within the vesicle population, as confirmed by cryogenic transmission electron microscopy (Cryo-TEM, Figure \ref{fig2}c). 

Temperature-dependent SAXS measurements revealed that vesicles containing $\SI{40}{\percent}$ cholesterol maintained structural symmetry at $\SI{45}\celsius$ \cite{BEATTIE20051760}.
This finding is consistent with the expected role of cholesterol in stiffening the membrane and reducing fluidity \cite{nagle2000structure}. 
Cryo-TEM imaging confirmed the presence of predominantly uni- and multi-lamellar, well-defined vesicles with diameters ranging from $\SI {200}{\nano\meter}$ to $\SI {300}{\nano\meter}$. No morphological differences attributable to cholesterol incorporation were observed (Figure \ref{fig2}c).
The membrane thicknesses of the photonic nanostructures determined from the work of Ghiradella were determined to be $\SI{24}{\nano\meter}$.\cite{Ghiradella1989}.
We found that incorporating up to 40 \% cholesterol into the lipid bilayers, Figure \ref{fig2}b, did not affect its thickness sufficiently to get to the thicknesses of $\approx \SI{24}{\nano\meter}$. 

Therefore, our results suggest that adding cholesterol to the lipidic membranes alone does not increase their thickness beyond  24\,nm, as observed in butterfly pupae \cite{Ghiradella1989}.
However, a substantial body of literature indicates that cholesterol plays an vital role in membrane fluidity or stiffness \cite{nagle2000structure}, and consequently in the stability of inverse bicontinuous LLLC mesophases.

\begin{figure}
\centering
\includegraphics[width=1\linewidth]{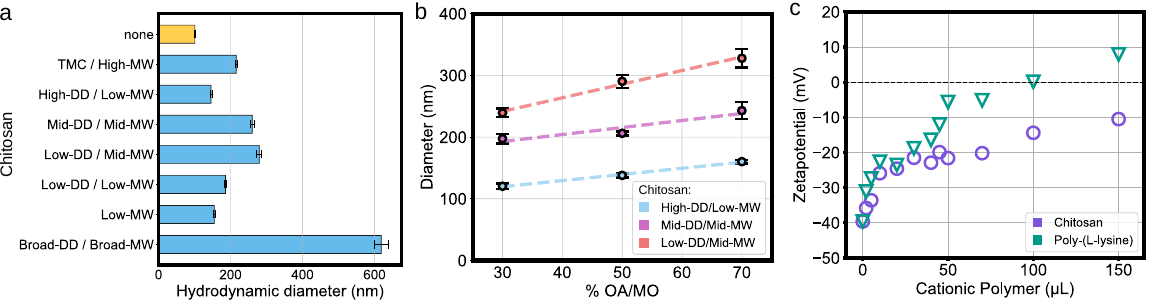}
\caption{\textbf{Influence of poly-cationic polymers on structure parameters of lipidic vesicles}. 
DLS and zetapotential measurements of lyotropic vesicles containing biopolymers.
(a) Hydrodynamic diameter of $\SI{30}{\percent}$ OA/MO vesicles extruded through a membrane with a 100\,nm pore size, with and without the addition of $\SI{0.1}{\percent}$ (w/v) chitosan with varying degrees of deacetylation (DD) and molecular weight (MW), in 30\,mM PBS and 100 mM\,NaCl at pH 7.0. This includes low-, mid-, and high DD of 70\%, 80\%, and 95\%, respectively, and low-, mid-, and high MW of 15-50\,kDa, 50-100 kDa, and 150-190 kDa, respectively.
(b) Hydrodynamic diameter and OA/MO ratio of vesicle extruded through a membrane with 100\,nm pore size after the addition of  $\SI{1}{\percent}$ (w/v) chitosan with varying DD and MW. 
(c) Zetapotential measurements comparing the charge neutralisation using $\SI{0.1}{\percent}$ broad-DD, broad-MW chitosan and $\SI{0.1}{\percent}$ poly(L-lysine) in 10\,mM PBS buffer for $\SI{30}{\percent}$ OA/MO vesicles.
}
\label{fig3}
\end{figure} 

\subsection{Biopolymers can restructure the lipid membrane}

Lyotropic lipid vesicles coated with biopolymer were also characterised.  
DLS and Zetapotential measurements were performed to investigate the effects of the polymer coating, the biopolymer encapsulation and the surface charge on the vesicles.
As illustrated in Figure \ref{fig1}, the two main types of biopolymer relevant to the explored parameter space are chitosan and poly(L-lysine). 
The surface charge of the lipid vesicles used in our methodology is primarily negative due the polar head-group of the oleic acid, which has an effective pKa of around 8 \cite{Salentinig2010}. 
The chosen biopolymers are cationic and were investigated for their ability to coat the negatively charged vesicles via electrostatic interactions.

\subsubsection{Effect of chitosan on vesicle size}
Figure~\ref{fig3}a,b show that adding the biopolymer chitosan to the lipid dispersion causes it to coat the lipid vesicles, increasing their overall diameter. 
This was measured by DLS at $\SI{90}{\degree}$. 
The hydrodynamic diameter and polydispersity index were derived from the peak intensity of the distribution and an average of three replicates.
Figure \ref{fig3}a shows the effect of different types of chitosan, including trimethyl chitosan (TMC), as well as chitosan with different degrees of deacetylation (DD), including $\SI{95}{\percent}$, $\SI{80}{\percent}$, and $\SI{70}{\percent}$. They are categorised as low-, mid-, and high DD of 70\%, 80\% and 95\%, respecively, and low-, mid- and high MW of 15-50\,kDa , 50-100 kDa and 150-190 kDa, for the different viscosities of chitosan, see Supplementary Table 1. 

Based on the previously reported phase diagram for oleic acid and monoolein \cite{Salentinig2010}, lipid vesicles can be produced at different OA/MO ratios ranging from $\SI{30}{\percent}$ to $\SI{100}{\percent}$ OA in an aqueous solution with a pH above 8.  Therefore, we have investigated three OA/MO ratios: ($\SI{30}{\percent}$, $\SI{50}{\percent}$ and $\SI{70}{\percent}$). 
The degree of deacetylation of the chitosan indicates the amount of positive charge available for protonation due to the primary amine on the polysaccharide subunit, D-glucosamine.
The samples were diluted in a pH 7 buffer, which deprotonates most of the chitosan. This suggests that the changes in vesicle diameter shown in Figure \ref{fig3}a , b are primarily driven by deacetylation degree, and secondarily by molecular weight.

As can be seen in Figure \ref{fig3}a, the size and polydispersity of the vesicles change when a $\SI{0.1}{\percent}$ (w/v) solution of chitosan is added, regardless of the deacetylation percentage, the molecular weight, or the oleic acid to monoolein ratio.
The highest hydrodynamic diameter was observed for vesicles coated with chitosan of a broad  molecular weight distribution.
Figure~\ref{fig3}b shows that the vesicle diameter depends on the deacetylation percentage and molecular weight of the chitosan. 
The largest difference in the hydrodynamic diameter of the chitosan-coated vesicles of 150\,nm occurs when  a $\SI{1}{\percent}$ (w/v) solution of $\SI{70}{\percent}$ deacetylated, high-molecular weight chitosan is added.
Adding a $\SI{1}{\percent}$ (w/v) solution of the same type of chitosan with $\SI{80}{\percent}$ deacetylation to different lipid dispersions results in a similar increase in diameter. This forms coated vesicles of approximately 600\,nm, as can be seen in Figure~\ref{fig3}a. This represents a 6-fold increase in diameter compared to the original diameter of 100\,nm for the uncoated vesicles.

This increase is due to the biopolymer coating as well as the water diffusing into the vesicles due to dilution caused by the added polymer solution and buffer to the vesicle dispersion.
The fact that dilution was equivalent across all chitosan-to-lipid ratios enables the comparison of the effect of chitosan coating on different vesicles. 
Additionally, the diameter of the vesicles appears to scale with the amount of oleic acid added for a given amount of chitosan, as shown in Figure~\ref{fig3}b.

\subsubsection{Effect of chitosan and poly-(L-lysine) on vesicle charge}
As can be seen from the zeta potential values in Figure~\ref{fig3}c,  the surface charge of the vesicles changes when they are coated with chitosan and poly(L-lysine).
Poly(L-lysine) was chosen because it becomes protonated at high pH.\cite{Stagi2022}
As more of the polymer is added, the surface charge reverses from negative to positive. This suggests that the vesicles are coated with the positively charged polymer, which neutralises their previous negative charge. 
The sharp initial decrease in the negative charge and subsequent tapering off for poly(L-lysine) indicates that the coating process is slower once \SI{50}{\micro\liter} of $\SI{0.1}{\percent}$ poly(L-lysine) solution has been added to the vesicle suspension.
Similar results for chitosan also suggest that the surface charge of the vesicles can be neutralised sufficiently, leading to aggregation of the vesicles in less dilute solutions  (see Supplementary Figure S2). 
The refractive indices of the two polymers
(chitosan: $n(\lambda=\SI{500}{\nano\meter})=1.522$ \cite{Azofeifa2012}, poly(L-lysine): $n(\lambda=\SI{589}{\nano\meter})=1.465$ \cite{Zhao2011}) 
are higher than that of the lipid vesicles (oleic acid, $n(\lambda=\SI{670}{\nano\meter})=1.466$ \cite{jones2015atmospherically}, monoolein, $n = 1.462$ \cite{Weast1979}).
Consequently, the effective refractive index of the coated vesicles is higher than that of the uncoated vesicles.

\begin{figure}
\centering
\includegraphics[width=0.75\linewidth]{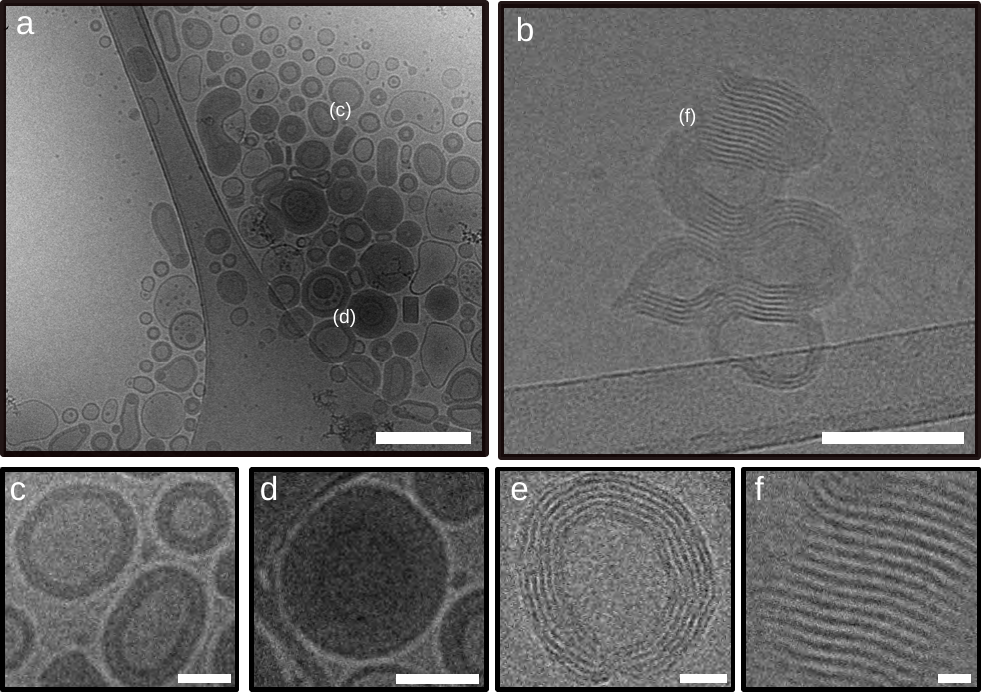}
\caption{
\textbf{Characterisation of Biomimetic Photonics Building Blocks.} 
Cryo-TEM images of lipid vesicles with poly-cationic biopolymers
(a) poly(L-lysine) (scale bar $\SI {1}{\micro\meter}$) and
(b) chitosan (scale bar 100 nm). Magnified images of vesicles 
(c) poly(L-lysine) coated and filled in intermembrane space (scale bar 100 nm), 
(d) poly(L-lysine) filled vesicle (scale bar 100 nm), 
(e) multi-lamellar vesicle with poly-cationic chitosan (low MW) (scale bar 50 nm), 
(f) vesicle disruption by poly-cationic chitosan (low MW) of lipid membrane and assembly of multi-lamellar bridged lipid membranes (scale bar 10 nm)}
\label{fig4}
\end{figure}

\subsubsection{Effect of polycationic polymers on lipid membrane stability}

Cryogenic TEM provides direct visual confirmation of the effect of poly(L-lysine) on lipid vesicles.
Figures \ref{fig4} a and b show how the addition of the cationic polymer destabilises the membrane of the lipid vesicles, resulting in membrane rearrangement and the adoption of a more structured geometry: Fig.~\ref{fig4}c coating; Fig.~\ref{fig4}d encapsulation; Fig.~\ref{fig4}e and f multi-lamellar stacks. 
The poly‑(L-lysine) coated vesicles from the same sample composition in as in  Fig.~\ref{fig4}c and d resemble standard lipid vesicles but have a thicker membrane coating of around 20 nm. 

Figure~\ref{fig4}c-f shows the vesicles in both separated and aggregated states. This suggests that the polymer coating triggers uncontrolled fusion and fission in some vesicles, while stabilising others. 
A uniformly spherical entity with a clearly defined, electron‑dense interior can also be seen. This is consistent with poly‑lysine encapsulation, either between the bilayer (Fig.~\ref{fig4}c) or within the lipid vesicles (Fig.~\ref{fig4}d). 

The largest observed vesicle has a diameter of approximately 400 nm in diameter, consistent with the order of magnitude obtained from the USAXS results, see Supplementary Fig S5. 
In contrast, excess poly-cationic polymers, such as low-MW chitosan, lead to membrane disruption and the formation of multilamellar structures. (as see Figure~\ref{fig4}e,f). 
The formation of concentric and asymmetric lamellae indicates the stacking of lipid membrane with alternating layers of cationic polymer. This morphological change suggests a concentration‑dependent transition from the vesicle phase to a multilayered assembly, wherein electrostatic bridging promotes lamellar stacking. 

Combined observations from X-ray scattering and electron microscopy demonstrate that cationic biopolymers can control the size and membrane thickness of lipid-biopolymer assemblies.  The internal ordering of the formed assemblies can be tuned in a pH‑dependent manner in the presence of the polymer, with a rearrangement pathway observed from monodisperse vesicles to multilamellar assemblies. This informs and guides experiments for the construction of biomimetic photonic crystals.
Trends in lipid vesicle size and polymer properties, as determined by DLS, zeta potential and cryo-EM suggest that optimising the molecular composition could result in the formation of coated or encapsulated lyotropic lipid vesicles. These could serve as building blocks in biomimetic photonic materials.

\begin{figure}[tbp]
\centering
\includegraphics[width=\linewidth]{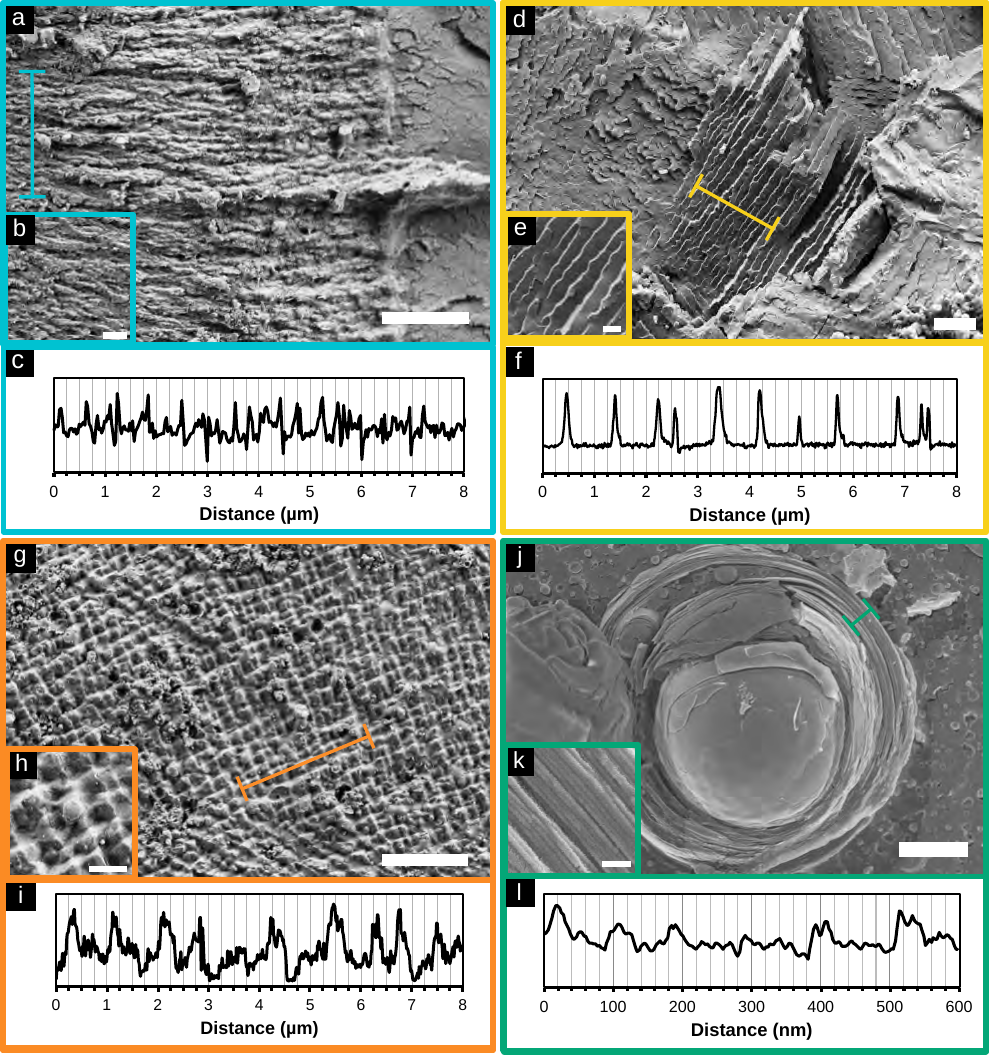}
\caption{\textbf{Periodic nanostructures from combined lipid/biopolymer systems -- one order of magnitude larger than lipidic systems}. Interaction between lipids and biopolymers, and reordering of lipid vesicles mixed with poly(L-lysine) and chitosan after 2 -10 days and vitrified using high pressure freezing. (a-c) 0.1\% chitosan and 0.1\% poly-(L-Lysine) solution mixture in the absence of lipids, (d-f) $\SI{30}{\percent}$ OA / $\SI{70}{\percent}$ MO and $\SI{0.1}{\percent}$ chitosan forming a corrugated pattern. (g-i) $\SI{35}{\percent}$ OA/MO with $\SI{33}{\percent}$ cholesterol and $\SI{0.1}{\percent}$ chitosan forming a cubic pattern. (j-l) $\SI{50}{\percent}$ OA/MO with $\SI{30}{\percent}$ cholesterol and $\SI{0.1}{\percent}$ Poly-(L-Lysine) forming concentric shells. The line graphs in(c,f,i,l) depict the height profiles and periodicities for each sample along the indicated lines in (a, d, g, and j). Scale bars: $\SI{5}{\micro\meter}$ for (a, d, g), $\SI{2}{\micro\meter}$ for (j), $\SI{1}{\micro\meter}$ for (b, e, h). $\SI{500}{\nano\meter}$ for (k)}
\label{fig5}
\end{figure}

\subsection{Lipids can affect polymer self-assembly at micro- and nanoscale}
The aggregation of vesicles under the influence of biopolymers can also be visualised over time by cryogenic scanning electron microscopy (Cryo-SEM).
Figure~\ref{fig5} shows Cryo-SEM images of the self‑assembled system containing lyotropic lipids chitosan and poly(L-lysine), which are the necessary molecular components. 
Images are taken of mixtures of lipid vesicles to which biopolymers were added and allowed to aggregate over periods ranging for 48 hours to 10 days. 
The samples were plunge-frozen in liquid nitrogen using a high pressure freezer, then fractured and sputter-coated prior to imaging. 
Figure~\ref{fig5}a, d and g show structures formed in solutions of chitosan and poly(L-lysine), either with or without OA/MO lipid vesicles.

Figure~\ref{fig5}b, e and i are magnified sections of the corresponding images in Figure~\ref{fig5}a,d and g.
Figure~\ref{fig5}c, f and h contain line graphs showing the periodicities observed in the sample. 

The first sample, depicted in Figure~\ref{fig5}a-c, contains only the polymer solution and no added lipids. As expected for frozen solutions of dissolved polymers with limited interaction, this sample exhibits no discernible periodicity. This can be seen in the line graph in Figure~\ref{fig5}c.
Figure~\ref{fig5}d and g depict samples containing both polymers and lipid vesicles. These samples are composed of $\SI{30}{\percent}$ OA and $\SI{70}{\percent}$ MO, and were extruded at a pH of 8.
Since chitosan dissolves in dilute acetic acid solution at pH levels below 5, adding  lipid vesicles to the biopolymer mixture increases the pH.
These samples exhibit emergent periodicity and appear to form a continuous polymer matrix containing  numerous lipid vesicles dispersed throughout the network, as seen in Supplementary Figure S4.
The vesicles were visible and spherical, with diameters ranging from 300 nm to 2 \textmu m (see Supplementary Figure S4).

The periodicities of the lipid-biopolymer mixture differ significantly from those of the biopolymer solution alone. 
Three main periodic nano- to microstructured patterns were observed in these samples: (i) corrugated (panels d-f) and (ii) cubic (panels g-i) for samples containing chitosan, and (iii) concentric shells for samples containing poly-(L-Lysine), (panels j-l).
The cubic domains had a periodicity ranging from 700 nm to 1.2 µm, the corrugated microstructure from 800 nm to 900 nm, and the concentric shells from 15 nm to 50 nm as measured from the line graphs in Figure~\ref{fig5}f,h, and i.
The cubic pattern can be interpreted as a cross-section of a three-dimensional structure resulting from freeze-fracture, likely due to the $Im\overline{3}m$ cubic mesophase of the oleic acid and monoolein lyotropic lipids present in the sample.
As the lipid content of the sample is diluted by the addition of the biopolymer, the visible periodic nanostructure is expected to comprise primarily chitosan and poly(L-lysine). 
Adding lipid vesicles to the biopolymer solution significantly increases the likelihood of observing periodicity, suggesting that the lipids may play a significant role in restructuring the polymer in solution.

\section{Discussion}

This study examines how molecular components known to be present in insects, particularly butterflies, can create periodic nanostructures with 100 nm length scales. The composition and concentration of each component is expected to influence the self-assembly process in a dynamic and complex solution, as depicted in Figure~\ref{fig1}. 

\subsection{Lipid composition and temperature alone are insufficient for producing photonic-scale structures}

The influence of oleic acid and monoolein lipid mixtures on the formation of selective mesophases in aqueous solutions was studied. A range of lipid solutions with a known oleic acid-to-monoolein ratio was selected using the previously reported phase diagram ~\cite{Salentinig2010} to ensure the presence of specific mesophases in solution. The vesicle phase was chosen for most of the reported work because changes in structure and size could be observed -- clear indicators that the self-assembled system was converging towards photonic dimensions. Previously published work suggests different possibilities about whether the vesicle phase precedes the gyroid phase.

\cite{Ghiradella1989, SeahSaranathan2023} 
To shed light on these different interpretations, we selected the vesicle phase. 
We found that although cholesterol is a ubiquitous component of insect lipid membranes and plays a definitive role in modulating membrane mechanics, it is insufficient for generating the periodicities required for visible-light structural colour on its own. This result is an important boundary condition for further experiments.

Cholesterol is a known component of lipid membranes in insects ~\cite{rog_cholesterol_2014}, particularly butterflies, and can insert into the bilayer, shifting the flexibility and curvature. 
It effectively increases membrane stiffness and reduces fluidity, as suggested by the preservation of structural symmetry in SAXS scans at elevated temperatures ($\SI{45}{\celsius}$) \cite{nagle2000structure}. 
This stiffening effect is consistent with the known role of cholesterol in ordering lipid tails and reducing the area per lipid molecule. 
However, the magnitude of the structural changes induced by cholesterol is limited to the nanometre scale. 
The observed increase in hydrodynamic diameter ($\approx 56\nm$) and the lack of significant change in bilayer thickness ($\approx 5\nm$) suggests that cholesterol primarily affects the local curvature and mechanical stability of the membrane. Our data show that stiffening the membranes with cholesterol does not increase the self-assembled dimensions to photonic length scales. 

The SAXS peaks at $\SI{13.4}{\nano\meter}$ and $\SI{24.2}{\nano\meter}$ correspond to multi-lamellar spacing rather than the thickness of a single bilayer. 
While these values are comparable to the $\SI{24}{\nano\meter}$ periodicities observed in butterfly pupae \cite{Ghiradella1989}, our results demonstrate that cholesterol alone cannot expand the lipid mesophase to the several hundred nanometres required for constructive interference in the visible wavelength range. 
The largest unit cells achieved in purely lipid systems, such as the electrostatically swollen cubic Im3m phase, typically do not exceed $\SI{68}{\nano\meter}$ \cite{Kim2017}. 
This discrepancy highlights a fundamental limitation of lipid-only self-assembly: without an external templating force or a secondary component to drive expansion, the thermodynamic minimum of the lipid system remains in the sub-visible regime.

Consequently, the role of cholesterol must be reinterpreted: rather than being a primary contributor to photonic periodicity, cholesterol stabilises intermediate mesophases. Its ability to rigidify the membrane may be essential for maintaining the structural integrity of lipid-biopolymer complexes during dynamic processes such as dehydration and reorganisation. In the context of natural insect development, cholesterol likely locks in the curvature induced by other factors, such as protein scaffolding or confinement, rather than generating that curvature de novo. This finding necessitates the inclusion of biopolymers in the experimental system to provide the additional driving force required to bridge the gap between nanometre-scale lipid ordering and 100\,nm-scale photonic structures.

\subsection{Biopolymers can guide lipid self-assembly via intermolecular interactions}

To understand the intermolecular interactions within the lipid membrane and the biopolymers involved in the self-assembly of photonic crystals in butterflies, we introduced poly-cationic biopolymers into the LLLCs system. 
The surface of the lyotropic lipid vesicles is negatively charged due to the presence of oleic acid at pH 8. 
We hypothesised that the positive charges of the polymers would cause the negatively charged lipid vesicles to interact and aggregate. Additionally, adsorption of the biopolymers onto the vesicle surface could reduce membrane instability and stabilise curvature, thereby encouraging the surrounding vesicles to align in periodic arrangements.

Incubation with the chitosan resulted in a significant increase in diameter (see Figure~\ref{fig3}a), indicating that the vesicles were coated with cationic chitosan. There was no significant difference in vesicle size depending on the lyotropic lipid mixture, which ranged from $\SI{30}{\percent}$ to $\SI{60}{\percent}$ oleic acid in the lipid solution with monoolein. 
Therefore, incorporating biopolymers into the lipid vesicle phase is a reliable method for investigating the types of biopolymer involved in the formation of natural photonic crystals.

Figure~\ref{fig3}b shows the effect of the molecular weight and deacetylation percentage of the chitosan in the experimental system, thus demonstrating the contribution of the positive charge. 
The greatest shift in diameter was observed for 600 nm vesicles incubated with high-molecular-weight chitosan with a $\SI{70}{\percent}$ cationic charge compared to lower-molecular-weight chitosan with $\SI{95}{\percent}$ cationic charge. Since chitosan is only positively charged at pH below its pKa of 5.5, most of the polymer is neutral at pH 8, when the lipids are in the vesicle phase. Figure~\ref{fig3}c shows the charge neutralisation effect using zeta potential measurements.  pH alters the ionisation state of lipid head‑groups, thereby modifying zeta potential. A poly(L-lysine) coating causes a more rapid change in the surface charge of the vesicles than a chitosan coating. 
This is an expected observation because poly(L-lysine) is positively charged throughout the pH range investigated in this study. In contrast, the addition of partially deacetylated chitosan results in a more gradual change in the surface charge.

Cryo-TEM data further demonstrate that poly(L-lysine) coats the surface of the vesicles and can also penetrate and be encapsulated within the lipid bilayer, thereby forming structured building blocks with diameters of approximately $\SI{500}{\nano\meter}$. 
The formation of multi-lamellar stacks and the potential transition to the HII phase upon the addition of chitosan indicates that the interaction between  the biopolymer and the lipid is pH- and concentration-dependent and capable of driving morphological transitions\cite{Spurlin2006}. 
The effective refractive index of the coated vesicles increases due to the higher indices of both chitosan ($n=1.522$) and poly(L-lysine) ($n=1.465$),  both of which are higher than the lipid core ($n=1.462$). This suggests that these biopolymer-lipid assemblies theoretically possess the necessary refractive index contrast with respect to air for photonic applications. However, with a polydispersity index of approximately 0.1, the 0.05 threshold required for a coherent photonic response of opal-like structures has not yet been reached \cite{Allard.2004}. This indicates that further optimisation of the coating and assembly process is required to achieve the monodispersity observed in natural photonic structures, such as those in \textit{Sternotomis mathildae} \cite{Bauernfeind2026}.

Taken together, these results reveal distinct structural trends for each biopolymer class. The proximity of chitosan to chitin is associated with an increase in the size of the periodic structure. In contrast, poly(L-lysine) promotes encapsulation and multi-lamellar stacking as shown in Figures ~\ref{fig4} and ~\ref{fig5} (j-l). 
These results provide the first indications of the formation of biomimetic photonic building blocks. It is hypothesised that the structure of these building blocks are responsible for the formation of photonic crystals in nature. These observations raise several questions for future research on this experimental system. For instance, which has a greater impact on the lipid membrane: ionic interactions or the steric effects of the biopolymer?

\subsection{The kinetics of self-assembly influence the formation of periodic structures}

Understanding how the lipid‑biopolymer system is affected by incubation time is essential for identifying how periodic nanostructures form in biological systems. 
Cryo‑SEM was used to observe how the lipid-polymer morphology arises from its interactions over extended time periods. 
Cryofixation preserves the instantaneous state of the vesicles and polymers in the dispersion, providing a direct view of the structures formed during self‑assembly. 
As all samples were processed the same way and fixed using high-pressure vitrification, any freezing-related effects  are expected to be consistent. 
Figure~\ref{fig5}a-c shows a relatively featureless polymer matrix composed of a $\SI{2}\percent$ chitosan and $\SI{0.1}\percent$ poly-(L-Lysine) solution. 
The absence of any regular pattern confirms that lipids are required for structure formation. 
Figure~\ref{fig5}d-f shows increasingly ordered structures: the addition of chitosan produces a corrugated morphology. 
The wave-like ridges appear to run primarily in one direction along the ordered sections, with a periodicity of $\SI{850}{\nano\meter}$. 
The peak symmetry results from the sharp contrast created by the corrugated edges and smooth sections of each ridge.

Similarly, in Figure~\ref{fig5}g-i demonstrates the impact of incorporating $\SI{30}\percent$ OA/MO vesicles into a solution of chitosan and poly(L‑lysine). This suggests that the mixture can self-assemble into a periodic network exhibiting a cross-hatch pattern. 
The orthogonal ridges visible in Figure~\ref{fig5}h suggest that a pattern resembling a 2D-cubic arrangement -- a cross-section through a three-dimensional structure -- can emerge from the lipid-polymer mixture. 
Since chitosan is only soluble at low pH,  adding this acidic solution to the vesicle suspension simultaneously reduces the pH of the solution. 
This cross-hatch pattern may therefore result from a mixture of the hexagonal HII phase of the LLC, which forms at a pH below 6.5, and the Im3m cubic phase, which forms at a pH of 6.75. 
The periodicity obtained from panel Figure~\ref{fig5}h was found to be $\SI{810}{\nano\meter}$, close to the largest size of chitosan coated vesicles observed by DLS  (see Figure~\ref{fig3}a).

One possible explanation for the periodic structuring of the corrugated assemblies and cubic domains is that it is due to multiple coated vesicles with internally ordered lipid mesophases merging over time to form larger domains spanning tens of micrometers\cite{TAKAHASHI199229}. We speculate that, in living organisms, these larger domains may then self-assemble to form photonic crystals. 
Further work with Cryo-FIB-SEM on such samples may be conducted in the future to confirm how far these patterns extend into a 3D arrangement. 
We hypothesise that if this value is close to the vesicle diameter, this would suggest that the vesicles act as a repeating unit. Conversely, shorter spacing could indicate the presence of polymer bridges between the lipid membrane, as observed in the Cryo-TEM analysis. Future work will systematically investigate the effect of changes in periodicity of formed microstructures over time on pH, ionic strength and biopolymer concentration.

\subsection{Lipid-biopolymer mixtures can produce periodic nanostructures on the micrometre scale}

Our analysis shows that neither cholesterol nor biopolymers can generate the necessary length scales for structural colour within the visible wavelength range alone. However,when they interact, they drive the formation of periodic nanostructures. Cholesterol provides the necessary membrane rigidity to stabilise intermediate mesophases, while cationic biopolymers such as chitosan and poly(L-lysine) induce electrostatic restructuring and may be encapsulated in lipid vesicles. 
Allowing these components to aggregate over time causes the system to transcend the sub-100 nm periodicity typical of pure lyotropic liquid crystals, resulting in ordered microstructures with periodicities ranging from $\SI{700}{\nano\meter}$ to $\SI{1.2}{\micro\meter}$. This emergent periodicity can be visualised as corrugated or cubic domains, as well as concentric shell patterns using Cryo-SEM. 
This confirms that the lipid vesicles act as templating units that reorganise the surrounding polymer matrix into a periodic stucture capable of interacting with visible light once the system has dried.

The observation that the chitosan analogue closest to chitin yields the largest periodic structures suggests that the living organisms likely exploit similar steric and ionic interactions to achieve their optical properties.  Furthermore, the ability to tune these periodicities by varying lipid composition, polymer charge density and incubation time provides a perspective for deconvoluting the complex developmental pathways of insect photonic crystals. 

\section{Conclusions}

We conducted experiments investigating the formation of structures in lipid and biopolymer mixtures that could contribute to the production of structural colour in insects. 
We observed unilamellar and multilamellar vesicles with diameters ranging from $\SI{100}{\nano\meter}$ to $\SI{600}{\nano\meter}$ and a polydispersity index close to 0.1. Our results confirm that, while cholesterol alone cannot generate photonic-scale periodicities, the combination of lyotropic lipid vesicles of approximately $\SI{100}{\nano\meter}$ and polycationic biopolymers enables the formation of periodic structures from $\SI{700}{\nano\meter}$ to $\SI{1.2}{\micro\meter}$ through self-assembly. 
This demonstrates the synergistic effect of lipids restructuring polymer networks and vice versa, creating a bioinspired methodology capable of reaching dimensions of 100 nm length scales up to even micrometers.

Bilayer spacings of $\SI{13}{\nano\meter}$ and $\SI{24}{\nano\meter}$, as observed in SAXS, also support a new hypothesis concerning the function of lipid membranes. It is hypothesised that membranes are multilayered because the interstitial space of multilamellar vesicles is filled with biopolymer due to its association with the lipid membrane. 
This hypothesis mechanistically explains the periodicities observed in TEM cross-sections of butterfly pupae by Ghiradella \cite{Ghiradella1989} and suggests that chitin accumulation within the interstitial spaces of multilamellar vesicles forms these structures rather than simple membrane stacking. The work reported in this manuscript indicates that the association of biopolymers and the lipid membrane is driven by selective intermolecular interactions, such as electrostatic attraction and hydrogen bonding. This mechanism corroborates the broader observation in the field of biophotonic nanostructures \cite{Saranathan2015} that various arthropod lineages have evolved to exploit the self-assembly of lipid-bilayer membranes to access a wide range of amphiphilic morphologies -- from bicontinuous networks to close-packed spheres -- and generate biophotonic nanostructures at optical length scales.

While the assembled networks exhibit the geometric prerequisites for photonic behaviour, no measurable photonic response was observed. This is likely due to a combination of incomplete ordering causing incoherent scattering and a low refractive index contrast in the aqueous system. 
Developing a solidification protocol that preserves long‑range periodicity is essential for creating nanostructures that could exhibit a photonic response. 
Future investigations into expanding our experimental system could explore alternative biopolymers, cross‑linking agents and controlled drying protocols to minimise defects.

The work provides a foundation for future investigations into: (i) how structural colour is produced in nature; (ii) which process parameters are worth investigating for producing biomimetic photonic materials; and (iii) the types of characterisation required to optimise the production of such biomaterials. 
A systematic experimental study involving the variation of parameters such as pH, polymer charge density, and kinetics is necessary. This will enable the parameter space of biomimetic photonic materials to be mapped, providing the field of soft matter with a fundamental understanding of the relationship between molecular self-assembly and photonic dimensions. Ultimately, it may be possible to establish a causal link between molecular composition, self-assembly parameters and the emergence of structural colour.  This could provide valuable insight for the rational design of tunable biomimetic photonic materials.

\section{Experimental Methods}
\subsection{Materials} \textbf{Lipids and Cholesterol:} Oleic acid (OA), monoolein (MO), and cholesterol were obtained from Sigma-Aldrich (Cat. No. 364525, 49960). 
All lipids were used as received without further purification. \ 
\textbf{Biopolymers:} Chitosan:
Chitosan 70/10 (Chitoscience GmbH, product no. 23201; degree of deacetylation \SIrange{67.6}{72.5}{\percent}; viscosity \SIrange{8}{15}{\milli\pascal\second}.
Chitosan 70/5 (Chitoscience GmbH, product no. 23200; degree of deacetylation \SIrange{67.6}{72.5}{\percent}; viscosity $\leq$\SI{7}{\milli\pascal\second}.
Chitosan 80/10 (Chitoscience GmbH, product no. 23401; degree of deacetylation \SIrange{77.6}{82.5}{\percent}; viscosity \SIrange{8}{15}{\milli\pascal\second}. All chitoscience viscosities reported for solutions of \SI{1}{\percent} (w/v) in \SI{1}{\percent} acetic acid at \SI{20}{\celsius}).
Chitosan (MW \SI{30}{\kilo\dalton}) (MedChemExpress, cat. no. HY-B2144B).
Trimethyl chitosan (MedChemExpress, Cat. No. HY-148033) with an average molecular weight of \SI{150}{\kilo\dalton}.
Chitosan (Sigma‑Aldrich, Cat. No. 448869) with molecular weight of \SIrange{50}{190}{\kilo\dalton}.
Poly(L-lysine) (\SI{0.1}{\percent} w/v in H$_2$O) was purchased from Sigma‑Aldrich (Catalog P8920‑100ML) with molecular weight of \SIrange{150}{300}{\kilo\dalton}. \ 
\textbf{Buffers and Reagents:} Phosphate-buffered saline (PBS) components (Na2HPO4, NaH2PO4, NaCl), sodium hydroxide (NaOH), hydrochloric acid (HCl), and ethanol/chloroform were of analytical grade and used without further purification. Pluronic F127 was sourced from Sigma-Aldrich (Catalog P2443). \ 

\subsection{Experimental Methods}

\textbf{Lipid Extrusion:} Lyotropic lipid vesicles were prepared following established protocols \cite{Hope1988}. 
Briefly, for vesicles containing $\SI{30}{\percent}$ oleic acid, $\SI{70}{\percent}$ monoolein, and $\SI{33}{\percent}$ cholesterol: $\SI{70}{\milli\gram}$ of monoolein, $\SI{30}{\milli\gram}$ of oleic acid, and $\SI{50}{\milli\gram}$ of cholesterol were weighed and dissolved in $\SI{2}{\milli\liter}$ of ethanol (or chloroform). 
The organic solvent was evaporated under vacuum with heating to $\SI{40}{\celsius}$ to form a thin lipid film. 
The film was hydrated with $\SI{1}{\milli\liter}$ of warmed sterile buffer ($\SI{30}{\milli\Molar}$ Phosphate Buffer, $\SI{100}{\milli\Molar}$ NaCl, $\SI{1}{\percent}$ F127 w/v, pH 8.1) and stirred at up to $\SI{45}{\celsius}$. 
The pH of the lipid dispersion was measured and adjusted if necessary. 
The dispersion was extruded 11 times through a nuclear track-etched membrane with a pore size of $\SI{0.6}{\micro\meter}$ using a hand-held extruder pre-heated to $\SI{45}{\celsius}$ at a flow rate of $\SI{1}{\milli\liter\per\minute}$. 
The resulting lipid vesicles were stored at $\SI{4}{\celsius}$ in a sealed vial. \ 
\textbf{Lipid-Biopolymer Experiments}: Lipid vesicles were aged for $\SI{24}{\hour}$ prior to biopolymer coating. 
The pH of the vesicle suspension was checked and adjusted to the target value using $\SI{1}{\Molar}$ HCl or $\SI{1}{\Molar}$ NaOH. 
Aliquots of $\SIrange{5}{160}{\micro\liter}$ of $\SI{0.1}{\percent}$ (w/v) poly(L-lysine) or chitosan solution were added to $\SI{100}{\micro\liter}$ of the lipid solution, bringing the total buffer volume to $\SI{1}{\milli\liter}$. 
The mixtures were incubated at room temperature for $\SIrange{30}{1440}{\minute}$ prior to characterization. 
For titration experiments, poly(L-lysine) ($\SI{24}{\micro\Molar}$) was added in aliquots to vesicle suspensions in buffer with pre-adjusted pH.

\subsection{Characterisation Methods}

\textbf{Dynamic Light Scattering (DLS):} Hydrodynamic diameter and polydispersity index (PDI) were measured using backscatter detection on an Anton Paar DLS instrument (Litesizer 501).
Measurements were performed at $\SI{25}{\celsius}$ with three replicates. 
\textbf{Zetapotential:} Zeta potential was determined by performing 20 scans up to $\SI{100}{\volt}$ at $\SI{25}{\celsius}$. 
Data were analyzed using the Smoluchowski approximation with a Henry factor of 1.5. 
\textbf{Cryo-Transmission Electron Microscopy (Cryo-TEM):} 
The samples were prepared in a controlled-environment vitrification system Vitrobot Mark~IV (Thermo Fisher Scientific (TFS), USA) at \SI{22}{\degree\Celsius} and \SI{100}{\percent} humidity. \SI{3.6}{\micro\litre} of sample was placed on hydrophylised lacey carbon-coated copper grids (EMS, USA), and the excess of sample was blotted with filter paper to form a thin film on the mesh holes. Then, the grids were plunge frozen into a mixture of liquid ethane/propane cooled by liquid nitrogen.

For automated loading in the cryo-TEM, the vitrified grids were clipped into AutoGrid sample carriers (TFS, USA). The data collection by cryo-TEM was carried out on a TFS Titan Krios (TFS, USA) operated at \SI{300}{\kilo\volt} of acceleration voltage equipped with a Falcon~III direct electron detector (TFS, USA). Digital images were recorded under low-dose conditions and an underfocus of \SI{1}{\micro\metre} to \SI{3}{\micro\metre} to get enough contrast.

\ 
\textbf{Cryo-Scanning Electron Microscopy (Cryo-SEM):} Small amounts of sample were carefully transferred with a pipette into \SI{6}{\milli\metre} aluminium planchettes and a second aluminium planchette was used as a lid. The planchette sandwiches were frozen in a high-pressure freezer HPM~100 (Bal-Tec/Leica, AT). Vitrified specimens were then transferred into a pre-cooled (\SI{-125}{\degree\Celsius}) freeze-fracturing system BAF~060 (Bal-Tec/Leica, AT) at \SI{1e-6}{\milli\bar} and fractured. Unidirectional tungsten deposition at an elevation angle of \SI{45}{\degree} to a thickness of \SI{3}{\nano\metre} was followed by \SI{3}{\nano\metre} at \SI{90}{\degree}. Transfer to the pre-cooled cryo-SEM was done under high vacuum (\SI{<5e-6}{\milli\bar}) with a cold air-lock shuttle VCT010 (Bal-Tec/Leica, AT). Cryo-SEM was performed in a field emission SEM (Merlin, Zeiss, DE) equipped with a cold stage to maintain the specimen temperature at \SI{-125}{\degree\Celsius} (VCT Cryostage, Bal-Tec/Leica, AT). In-lens SE- and Everhart--Thornley SE-signals at an acceleration voltage of \SI{2}{\kilo\volt} were used for image acquisition. The contrast and brightness of the pictures were adjusted if necessary. \ 
\textbf{X-ray Scattering (SAXS/USAXS):} Measurements were acquired at the European Synchrotron Research Facility (ESRF), Grenoble, France, on beamline ID02. Scans were conducted in quartz capillaries as line scans with an exposure time of \SI{0.5}{\second} and 10 spots along the capillary to minimize beam damage. The beam energy was 12 keV and with dimensions $\SI{130}{\micro\meter}$ by $\SI{40}{\micro\meter}$ the sample-to-detector distance was \SI{1}{\meter} for SAXS and \SI{10}{\meter} for USAXS.

\medskip
\textbf{Supporting Information} \par 
Supporting Information is available the corresponding authors upon request.

\medskip
\textbf{Acknowledgements} \par 
The authors acknowledge funding from the SNSF Spark grant under the number CRSK-2\_220805 and the SNSF Ambizione grant under the number PZ00-2\_233007. Both were awarded to Viola Vogler-Neuling. 
We thank the beamline scientists for help with experiments at the European Synchrotron Research facility at the ID02 beamline, namely Gouranga Manna under the grant proposal SC-5596 awarded to Andrea Dodero, and Austin Hubley under the grant SC-5723 awarded to Viola Vogler-Neuling and Rushna Quddus. We also thank the ScopeM facility at ETH Zurich for help with Cryo-TEM and Cryo-SEM imaging, especially the microscopists Stephan Handschin and Miroslav Peterek.  
We additionally thank Sandor Balog for help with SAXS and USAXS data analysis.
\medskip
\bibliographystyle{MSP}
\bibliography{Biophotonics_Quddus_Refs.bib}
\end{document}

\author{%
Rushna Quddus\textsuperscript{1,*} \quad
Ullrich Steiner\textsuperscript{1} \quad
Viola Vogler-Neuling\textsuperscript{1,*}\\[1em]
\textsuperscript{1}Adolphe Merkle Institute, Fribourg, Switzerland\\
\textsuperscript{*}Corresponding authors: rushna.quddus@unifr.ch; viola.vogler-neuling@unifr.ch%
}
\maketitle

\section{Physicochemical Basis of Chitosan Properties}

\subsection{Chitosan Degree of Deacetylation (DD) and Charge Density}
Chitosan is a linear polysaccharide composed of randomly distributed $\beta$-(1$\to$4)-linked D-glucosamine (deacetylated unit) and N-acetyl-D-glucosamine (acetylated unit). The Degree of Deacetylation (DD) is defined as the molar fraction of glucosamine units in the polymer chain.
The cationic nature of chitosan arises from the protonation of the primary amine groups ($-\text{NH}_2$) located at the C2 position of the glucosamine units. In aqueous solutions with pH below the pKa of the amine group (typically $\approx 6.5$), these groups undergo protonation:

\begin{equation}
\text{R}-\text{NH}_2 + \text{H}^+ \rightleftharpoons \text{R}-\text{NH}_3^+
\end{equation}

Consequently, the theoretical maximum charge density ($\sigma_{max}$) of a chitosan chain is directly proportional to its DD. A higher DD implies a greater number of available protonation sites per unit mass, resulting in:

\begin{enumerate}
    \item Increased electrostatic attraction to anionic species (e.g., oleic acid headgroups).
    \item Increased chain extension due to intramolecular electrostatic repulsion between adjacent protonated amines.
    \item Positive trend of zeta potential values for chitosan-coated vesicles at equivalent concentrations.
\end{enumerate}

Conversely, lower DD values (higher acetylation) reduce the net positive charge, potentially altering the binding stoichiometry and the stability of the resulting lipid-polymer complexes.
Trimethyl chitosan (TMC) contains permanent quaternary ammonium cations. Unlike standard chitosan, which requires acidic conditions (pH < pKa) for protonation, TMC maintains a positive charge across the entire relevant pH range. This effectively permanent ionization ensures that every glucosamine unit contributes to the net charge, rendering its effective cationic density functionally equivalent to chitosan with a 100\% degree of deacetylation.

\subsection{Chitosan Viscosity and Molecular Weight}
\begin{table}
\centering
\caption{Properties of Chitosan used in the study}
\begin{tabular}{lllll}

\toprule
\textbf{Name} & $\textbf{DD (\%)}$ & \textbf{Viscosity (mPas)} & \textbf{$M_v$ (kDa)} & \textbf{Source} \\

\midrule
High-DD Low-MW & 92.6 & $\leq$7 & $\sim$15--50 & Chi.Sci.\textsuperscript{a} \\
Mid-DD Mid-MW & 77.6--82.5 & 8--15 & $\sim$50--100 & Chi.Sci.\textsuperscript{a} \\
Low-DD Low-MW & 67.6--72.5 & $\leq$7 & $\sim$15--50 & Chi.Sci.\textsuperscript{a} \\
Low-DD Mid-MW & 67.6--72.5 & 8--15 & $\sim$50--100 & Chi.Sci.\textsuperscript{a} \\
Low-MW Standard & -- & $\sim$3--8 & 30 & MCE\textsuperscript{b} \\
Broad-DD Broad-MW & $\geq$75 & 20--300 & 50--190 & Sig.-Ald.\textsuperscript{c} \\
TMC High-MW & $\sim$100 & -- & 150 & MCE\textsuperscript{b} \\

\bottomrule
\end{tabular}
\begin{flushleft}
\footnotesize
\textsuperscript{a} Chitoscience \\
\textsuperscript{b} MedChemExpress \\
\textsuperscript{c} Sigma-Aldrich \\
$\sim$ for calculated estimates
\end{flushleft}
\label{SI_Table1}
\end{table}

The intrinsic viscosity of chitosan in dilute acid solutions is governed by the Mark-Houwink-Sakurada equation (Kasaai, 2007)\footnote{Kasaai, M.R. (2007) Calculation of Mark--Houwink--Sakurada (MHS) equation viscometric constants for chitosan in any solvent--temperature system. Carbohydr. Polym. 68, 477--488.},

\begin{equation}
    [\eta] = K M^\alpha
\end{equation}

where $M$ is the molecular weight and K and $\alpha$ depend on supplier-given DD.
While DD influences chain stiffness and solubility, the viscosity grade (e.g., 5 mPas vs. 10 mPas) reported by suppliers is primarily an indicator of the molecular weight distribution.
In the context of the study:

\begin{itemize}
    \item \textbf{High Viscosity (High MW):} Longer polymer chains provide greater steric stabilization and can bridge multiple lipid vesicles, potentially leading to larger aggregate sizes or gelation at lower concentrations.
    \item \textbf{Low Viscosity (Low MW):} Shorter chains may penetrate the lipid bilayer more readily or form thinner surface coatings, influencing the permeability and mechanical properties of the composite structure.
\end{itemize}

The interplay between DD (charge) and MW (chain length) dictates the final architecture of the self-assembled nanostructures.
The specific grades utilized in this study, specifically for Figure 2 characterized by their supplier-reported DD ranges and viscosity, are summarized below. Note that the molecular weight ($M_v$) is estimated based on the viscosity data using the Kasaai (2007) constants for the specific solvent system (1\% acetic acid). Degree of deacetylation (DD) cannot be calculated from viscosity or $M_v$ (independent property; supplier-provided where available). For entries with reported $M_v$, viscosity was back-calculated using the same relation.

\section{Supplementary Figures}
%\newpage
\begin{figure}[H]
\centering
\includegraphics[width=0.6\linewidth]{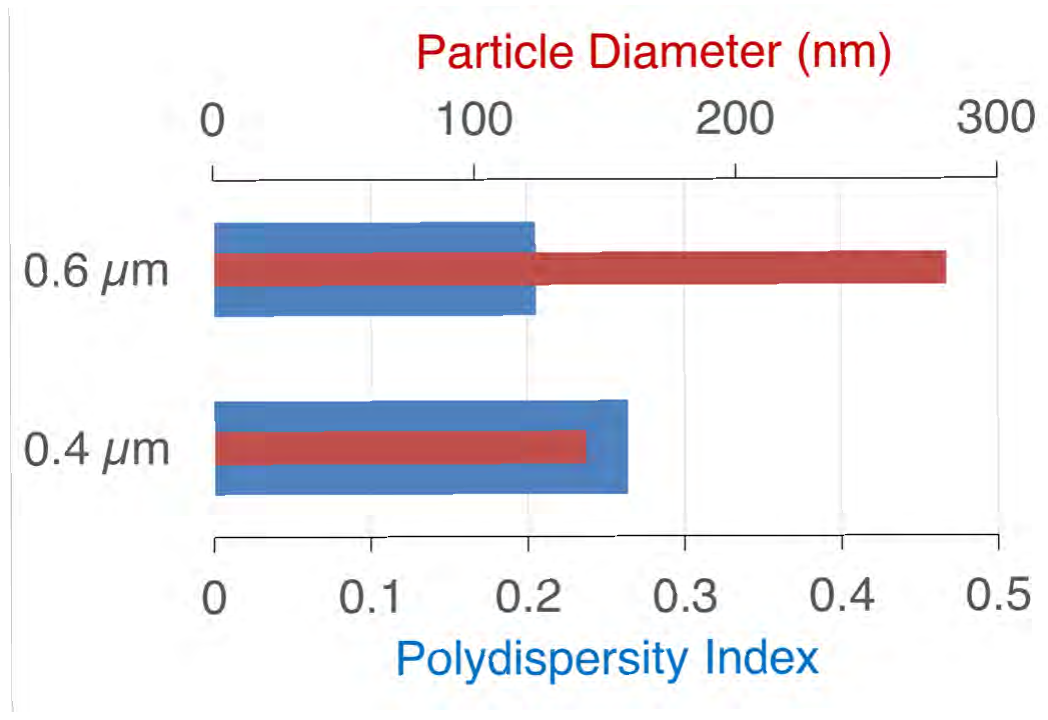}
\caption{\textbf{Effect of pore size on vesicles with Oleic acid and Monoolein.} Different pore sizes showed variation in polydispersity index but no significant change in vesicle hydrodynamic diameter. Initial lipid dispersion made by vortexing and sonication may already contain vesicles of approximately 150 nm diameter with no subsequent difference in using larger pore sizes. 
}
\label{S1}
\end{figure}  
%\clearpage

\begin{figure}
\centering
\includegraphics[width=0.7\linewidth]{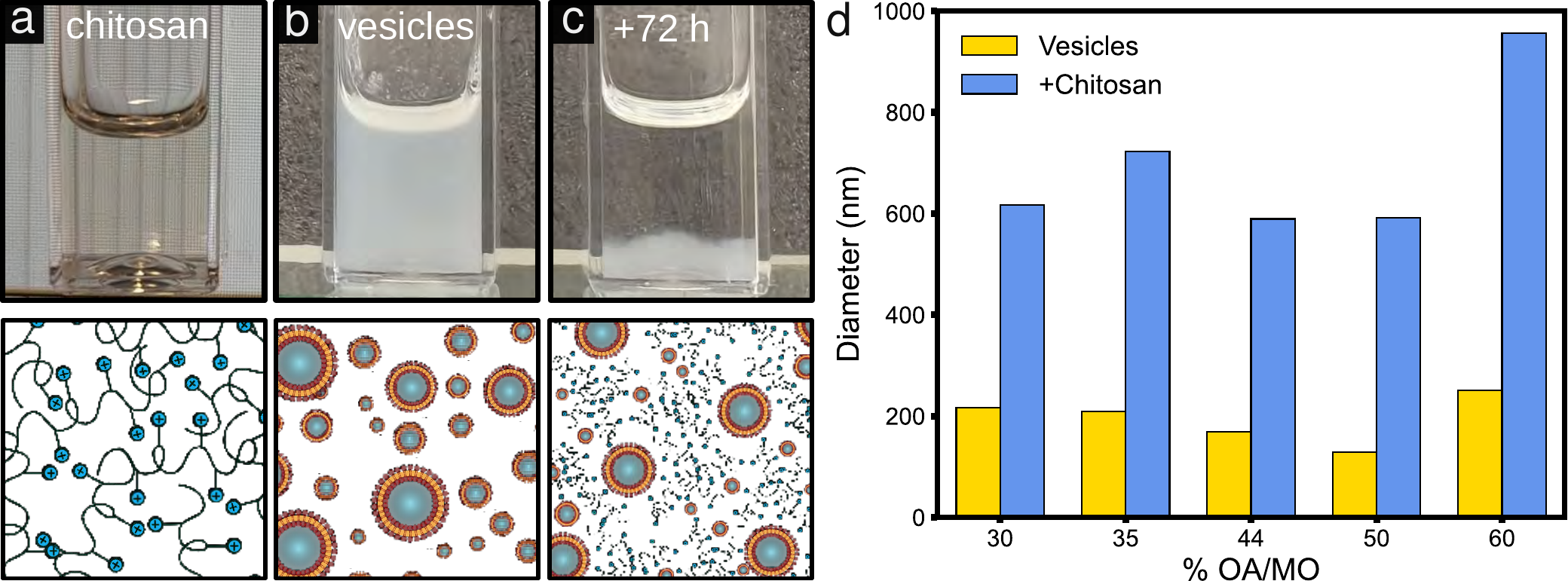}
\caption{\textbf{Aggregation of lipid vesicles with biopolymer.}
Photos and schematics of solutions of the poly-cationic polymer, chitosan (a), the oleic acid and monoolein lyotropic lipid vesicles in dispersion after extrusion and dilution with PBS buffer (b) and when coated with the chitosan, which can visibly scatter light eventually leads to aggregation or coacervation of the coated vesicle dispersion into a visible precipitate or dense gel (c) when stored for 3 days at 25◦ C.}
\label{S2}
\end{figure}
%\clearpage

\begin{figure}
\centering
\includegraphics[width=0.7\linewidth]{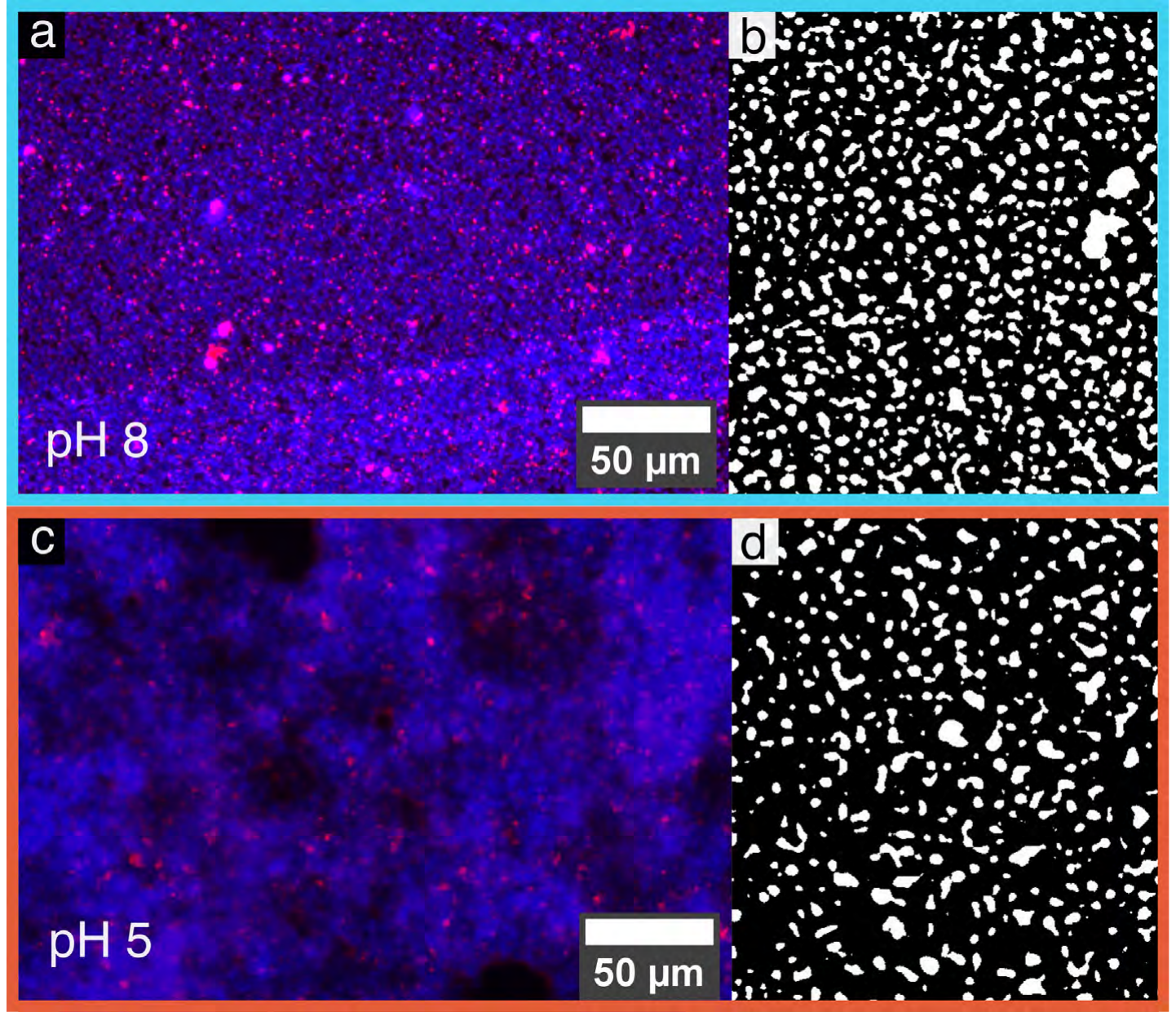}
\caption{\textbf{Quantitative Image Analysis of fluorencence microscopy of lipid-biopolymer network.}
Fluorescence microscopy of oleic acid and monolein lipid vesicles co-localised in chitosan and poly-(L-Lysine) bioplymer network at pH 8 (a, b) and pH 5 (c,d) after dehydration in humidity chamber with 90\% humidity over 48 hours. Blue fluorescence from Fluorescent brightner 28 dye for hydrophilic polymer phase and red fluorescence for lipid phase with Nile Blue A lipophilic dye. Quantitative image analysis performed using Ilastik and Fiji with segmentation and object identification to compare number of discrete lipid regions for both pH values.
}
\label{S3}
\end{figure}
\clearpage

\begin{figure}
\centering
\includegraphics[width=1\linewidth]{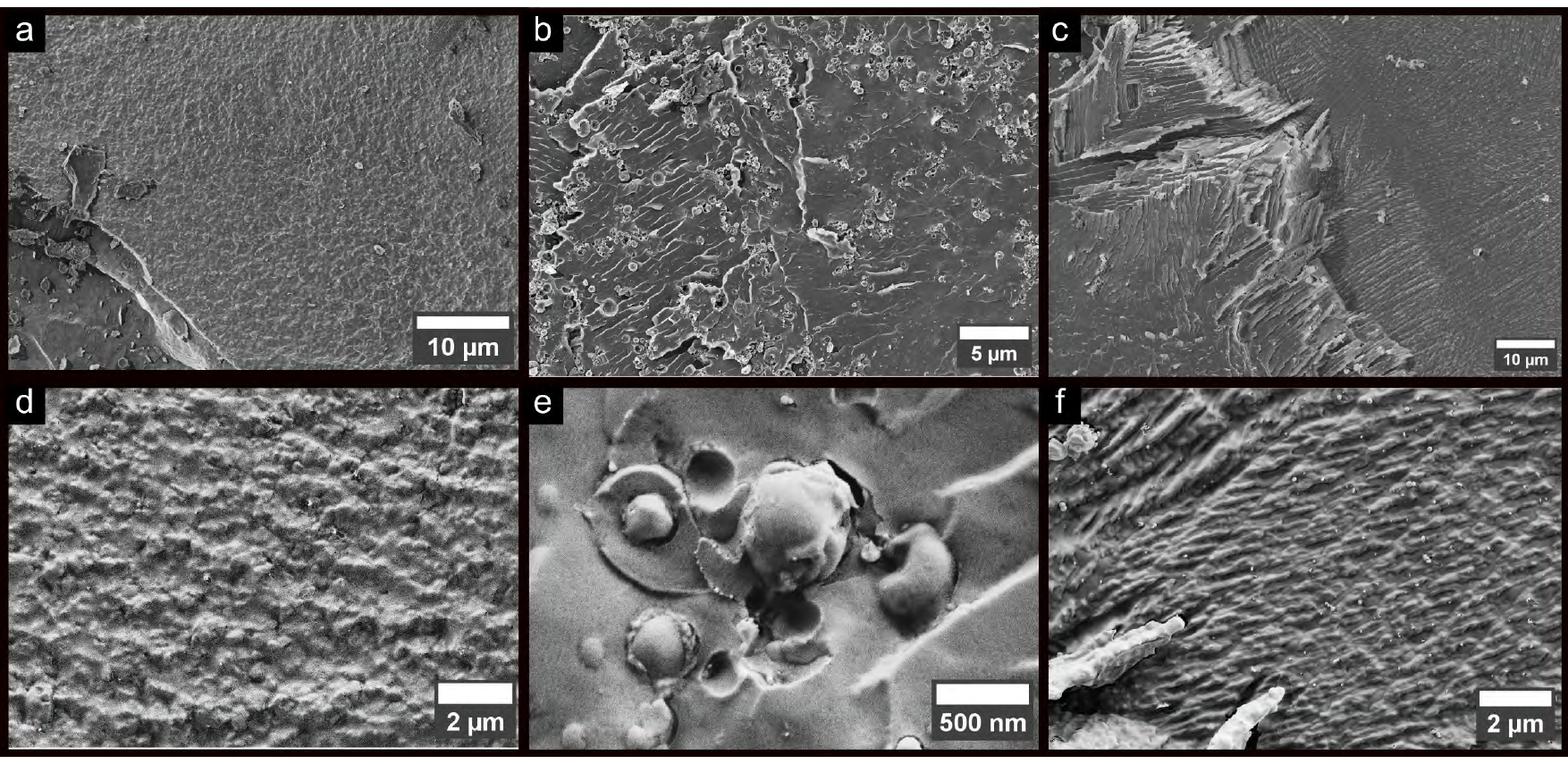}
\caption{\textbf{Cryo-SEM of lipid-biopolymer self-assembled system}
Panels a, d: Chitosan and poly-lysine solution showing large areas of sample with aperiodic surface, Panels b,e and c,f : 30\%/70\% Oleic acid and monoolein mixture with Chitosan and poly-lysine. Corrugated and partial cross-hatch patterns and embedded lipid vesicles can be distinguished
in different parts of the sample
}
\label{S4}
\end{figure}
%\clearpage

\begin{figure}
\centering
\includegraphics[width=1\linewidth]{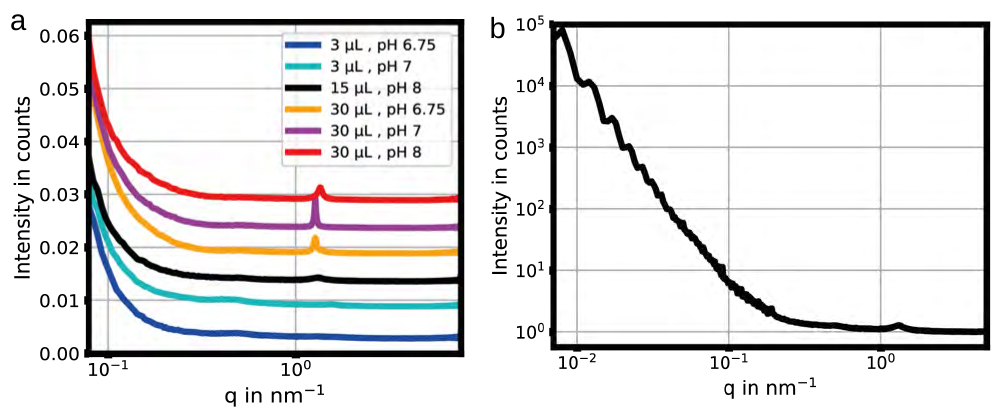}
\caption{
\textbf{Characterisation of vesicles with poly-(L-Lysine).} 
(a) SAXS analysis of $\SI{30}{\percent}$ OA/MO lipid vesicles with increasing amounts of a  24 \textmu M poly(L-lysine) solution added at various pH values, showing the effect of biopolymers on lipid self-assembly. 
(b) USAXS analysis qualitatively showing the formation of vesicles filled with poly(L-lysine).
}
\label{S5}
\end{figure}
%\clearpage

%%%%%%%%%%%%%%%%%%%%%%%%%%%%%%%%%%%%%%%%%%%%%%%%